\documentclass[conference]{IEEEtran}

\ifCLASSINFOpdf
\else
\fi

\usepackage{amsmath}
\usepackage{amssymb}
\usepackage{graphicx}
\usepackage[ruled,vlined]{algorithm2e}
\usepackage{multirow}
\usepackage{algorithmic}
\begin{document}

\title{PRVNet: A Novel Partially-Regularized Variational Autoencoders for Massive MIMO CSI Feedback}

\author{
\IEEEauthorblockN{Mostafa Hussien\textsuperscript{1,2}, Kim Khoa Nguyen\textsuperscript{1}, and Mohamed Cheriet\textsuperscript{1}}
\IEEEauthorblockA{\textsuperscript{1} École de technologie supérieure (ÉTS), Univeristy of Québec, Canada.\\
\textsuperscript{2} Information Technology Department, Assiut University, Egypt.}
}

\maketitle

\begin{abstract}
In a multiple-input multiple-output frequency-division duplexing (MIMO-FDD) system, the user equipment (UE) sends the downlink channel state information (CSI) to the base station to report link status. Due to the complexity of MIMO systems, the overhead incurred in sending this information negatively affects the system bandwidth. Although this problem has been widely considered in the literature, prior work generally assumes an ideal feedback channel. In this paper, we introduce PRVNet, a neural network architecture inspired by variational autoencoders (VAE) to compress the CSI matrix before sending it back to the base station under noisy channel conditions. Moreover, we propose a customized loss function that best suits the special characteristics of the problem being addressed. We also introduce an additional regularization hyperparameter for the learning objective, which is crucial for achieving competitive performance. In addition, we provide an efficient way to tune this hyperparameter using KL-annealing. Experimental results show the proposed model outperforms the benchmark models including two deep learning-based models in a noise-free feedback channel assumption. In addition, the proposed model achieves an outstanding performance under different noise levels for additive white Gaussian noise feedback channels. 
\end{abstract}


\IEEEpeerreviewmaketitle

\section{Introduction}
\label{sec:Introduction}

\begin{figure}[t]
\centerline{\includegraphics[width=0.49\textwidth]{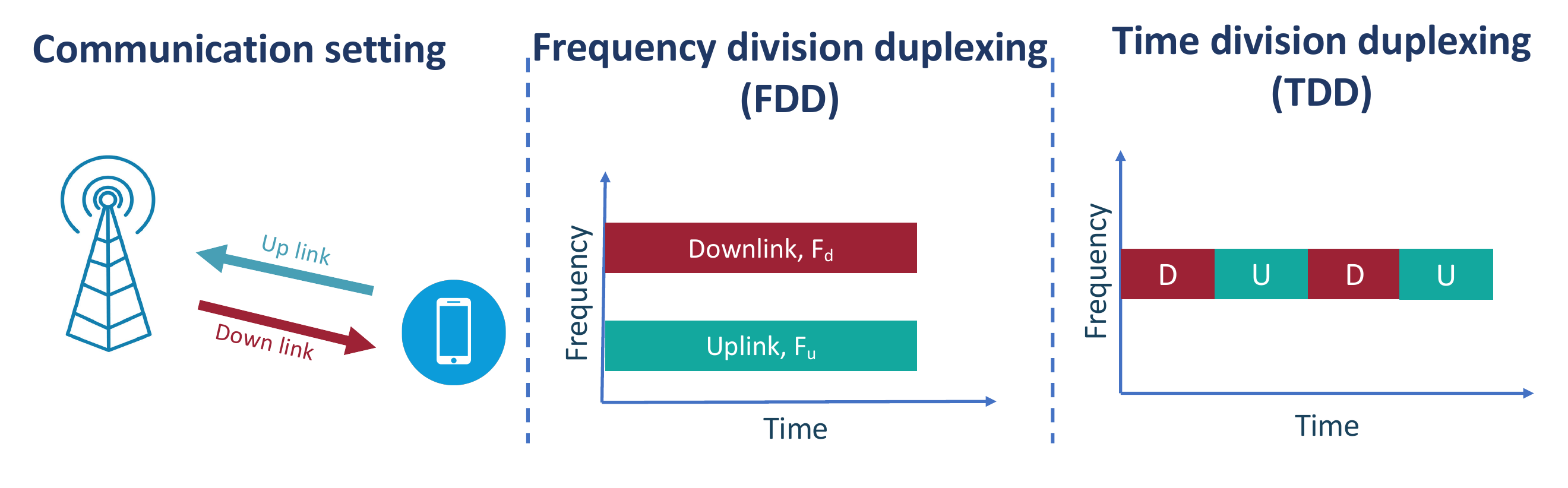}}
\caption{Channel reciprocity in TDD and FDD systems.}
\label{fig:TDD_FDD}
\vspace{-0.5cm}
\end{figure}

Multiple-input multiple-output (MIMO) system is considered as a key enabling technologies for fifth-generation, 5G, wireless systems. One of the active research areas in MIMO systems is channel state information (CSI) compression for feedback. In modern MIMO systems, a base station (BS) can be equipped by several antennas to reduce the multiuser interference and increase the cell throughput. In this setting, the BS is required to perform the precoding at its side. Therefore, the BS should have access to the current CSI. In time division duplexing (TDD) systems, the downlink CSI can be estimated using any channel estimation technique at the BS side. This is achievable because channel reciprocity holds for TDD systems, (Fig. \ref{fig:TDD_FDD}). However, in frequency division duplexing (FDD) systems, the uplink and downlink channels use different frequency bands, so channel reciprocity is no longer holds. In this case, the downlink CSI should be sent to the BS by the user equipment (UE). In modern MIMO systems, this channel matrix is huge and the bandwidth overhead incurred for sending this matrix can heavily degrades the system performance.

To alleviate this problem, the UE can compress the CSI matrix before sending it back to the BS. The compressed CSI matrix, however, should maintain enough information about the original CSI matrix in order for the BS to accurately reconstruct the original CSI matrix. Obtaining a low fidelity reconstruction for the original CSI matrix results in poor system performance. The problem then is how to optimally compress the CSI matrix while preserving its salient features and information. In the same time, the compression/decompression processes should be completed in real-time. Moreover, the encoder part should not consume much space or power since it resides at the UE which might have limited space and power resources. Lastly, and more importantly, this compressed data will be sent to the BS over a wireless channel (uplink channel) which suffers from traditional wireless transmission impairments such as noise, fading, or path loss. How the compression technique be resilient against the varying noise conditions of the uplink channel adds one more challenge to the CSI feedback problem. 

The problem of compressing CSI feedback has been considered in the literature. Traditional methods \cite{kuo2012compressive, lu2015sparsity} considered compressive sensing (CS) technique to compress the CSI matrix before feeding it back to the BS. However, the CSI matrix should maintain a high degree of sparsity for these methods to work. This condition is not always guaranteed in complex communication systems. In addition, many of these techniques depend on iterative approaches for solving a system of equations which makes them suffer a relatively slow performance.

On the other hand, artificial intelligence (AI) and deep learning (DL) have shown an outstanding performance in solving different complex problems in wireless communications \cite{wang2017deep, hussien2021towards}. A line of work has utilized AI and DL techniques to solve the CSI feedback problem. The authors in \cite{wen2018deep} opened the door for applying DL techniques in CSI feedback problem. They presented \textit{CsiNet}, a convolutional neural network architecture with skip connections in the decoder part. The advantage of CsiNet performance has been proven against traditional CS-based techniques. However, CsiNet employs a point estimation architecture in which the model learns one scalar value for each dimension in the codeword. This results in noise-sensitive codewords, and any noise level can largely hurt the reconstruction fidelity at the BS. Unlike CsiNet, our proposed model in this paper approximates distribution parameters for each dimension. In particular, mean and variance for a Gaussian latent space. This makes our codewords more robust against noises, and the decoder has the capacity to reconstruct the received codewords even in the presence of relatively large noise levels.

The authors in \cite{lu2019mimo} exploited the temporal and frequency correlations of wireless channels. They presented \textit{CSINet-LSTM}, which extends CsiNet with long short term memory (LSTM) network. LSTM is a classic type of recurrent neural network capable of capturing long-term dependencies (temporal correlation) between input sequences. In \cite{lu2020multi}, the authors proposed a neural network architecture, called \textit{CRNet}, for multi-resolution CSI feedback in massive MIMO. Their model achieves better performance than classical CS-based techniques as well as CsiNet. Another extension to CsiNet called \textit{CsiNet+} has been introduced in \cite{guo2020convolutional}. However, the floating point operations (flops) in CSINet+ is much larger than CsiNet, therefore the improvements come at the cost of complexity. 

In general, the limitations of prior work consists in: a) most of prior work assume an ideal control channel and pay less attention to the more practical assumption of noisy feedback channels, and b) no prior work has deeply investigated the power of generative models, especially variational autoencoders (VAE) \cite{kingma2013auto}, in the context of CSI compression despite their proven performance in many applications.

In this work, we propose a VAE-based framework for CSI feedback compression in MIMO-OFDM systems. The proposed framework customizes the VAE loss function to suit the special characteristics of the CSI feedback problem while, in the same time, benefiting from the robustness of the VAE-generated codewords against noise. The main contributions of this work can be summarized as follows:
\begin{itemize}
    \item A novel partially-regularized VAE model, named PRVNet, for CSI feedback problem with a new objective function that reflects the specific characteristics of CSI compression.
    \item A seminal algorithm inspired by  Kullback-Leibler (KL)-annealing to fine-tune the additional hyperparameter introduced in the objective function.
    \item We consider the CSI feedback in an additive white Gaussian noise (AWGN) channel. Since we employ a distribution-estimation model, our proposed model is shown to be capable of reconstructing the CSI matrices with high accuracy even under high noise levels.
\end{itemize}

The rest of this paper is organized as follows: in section \ref{sec:syst_model}, we present the system model. Section \ref{sec:proposed_model} presents a detailed description for the proposed PRVNet model, its architecture, and the training algorithm. The results of the proposed model along with comparisons against state-of-the-art works are presented in Section \ref{sec:sim_results}, followed by a conclusion. 

\section{System Model}
\label{sec:syst_model}

\begin{figure*}[t]
\centerline{\includegraphics[width=0.9\textwidth]{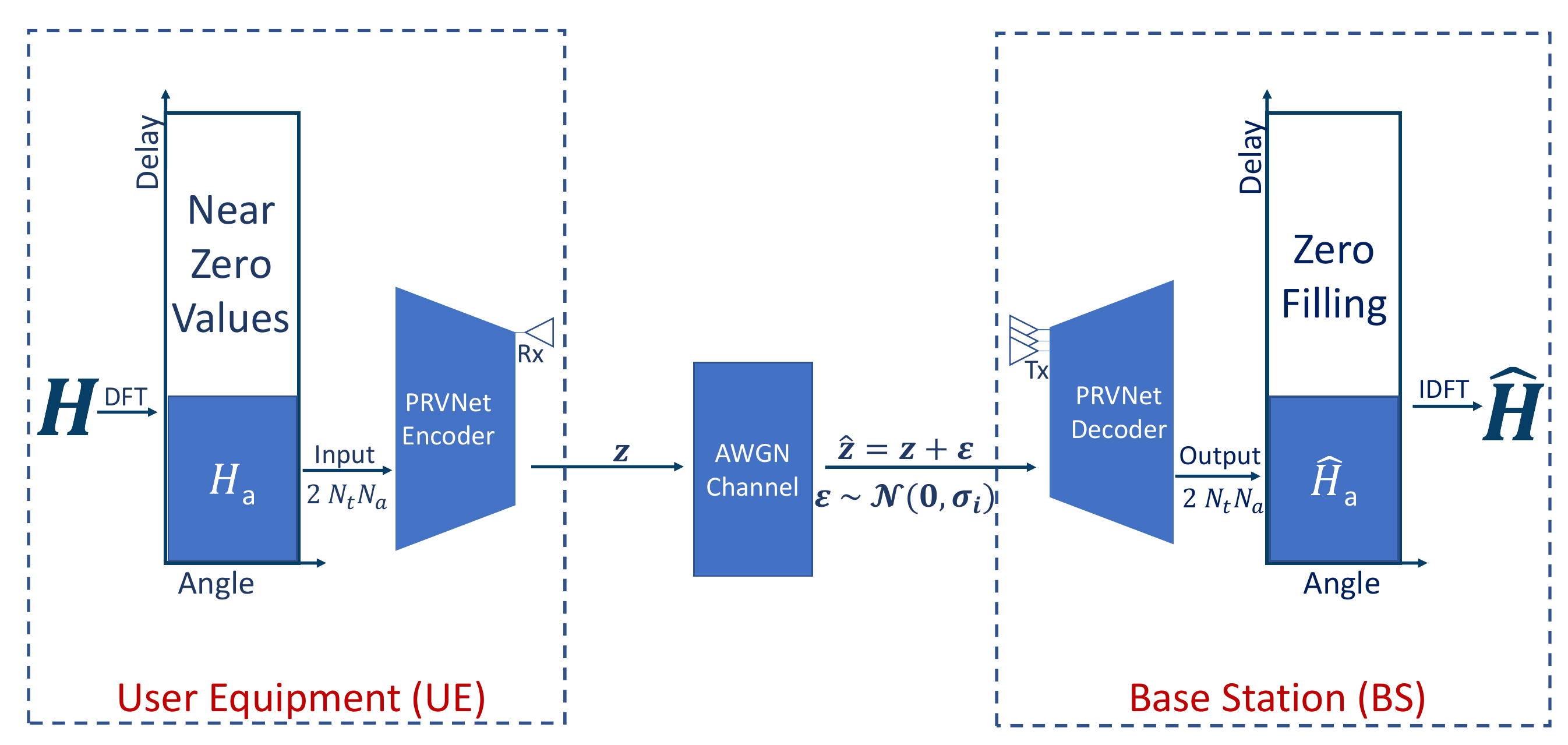}}
\caption{An overview of the PRVNet for CSI feedback compression through an AWGN channel.}
\label{fig:system_fig}
\end{figure*}

We consider a simple single-cell downlink massive MIMO system with $N_t\gg1$ transmit antennas at the BS and a single receive antenna at the UE. The system employs OFDM with $\Tilde{N}_c$ subcarriers. The received signal at the $n_{th}$ subcarrier, $y_n$, is given by:

\begin{equation}
    y_n = {\mathbf{\Tilde{h}}^ \mathbf{H}}_n \mathbf{v_n} x_n + z_n,
\end{equation}

where $ \mathbf{\Tilde{h}}_n \in \mathbb{C}^{N_t \times 1} $, $\mathbf{v}_n \in \mathbb{C}^{N_t \times 1}$, $ \mathit{x}_n \in \mathbb{C}$, and $z_n \in \mathbb{C}$ denote the channel vector, precoding vector, data-bearing symbol, and additive noise of the $n_{th}$ subcarrier, respectively. Also, assume $\mathbf{\Tilde{H}}$ = $\left [ \mathbf{\Tilde{h}}_1 \dots  \mathbf{\Tilde{h}}_{\Tilde{N}_c}  \right ] \in \mathbb{C}^{\Tilde{N}_c \times N_t}$ be the CSI stacked in the spatial frequency domain. The BS can design the precoding vectors $\{ \mathbf{v}_n, \; n=1, \dots \Tilde{N}_c\}$ once it receives $ \mathbf{\Tilde{H}}$ feedback. 

In FDD systems, the BS continually receives the channel matrix, $ \mathbf{\Tilde{H}}$, through a feedback links. This feedback has an $N_t \times \Tilde{N}_c$ dimensions. Estimating this channel at the UE side is out of the scope of this work. We assume that a perfect CSI has been acquired through pilot-based training \cite{choi2014downlink} and this work focuses on the feedback scheme.

To reduce the feedback overhead, we propose that $ \mathbf{\Tilde{H}}$ can be sparsified in the angular-delay domain using a 2D discrete Fourier transform (DFT) as follows:
\begin{equation}
\label{eq:discr_FT}
    \mathbf{H} = \mathbf{F_d }  \mathbf{\Tilde{H}} \mathbf{F^H_a},
\end{equation}
where $\mathbf{F_d}$ and $\mathbf{F^H_a}$ are $\Tilde{N}_c \times \Tilde{N}_c$ and $N_t \times N_t$ DFT matrices, respectively. Only a small fraction of the elements of $\mathbf{H}$ are large components, and the remainders are close to zero. In the delay domain, only the first $N_a$ rows of $\mathbf{H}$ contain values because the time delay between multipath arrivals lies within a limited period. Therefore, we can retain the first $N_a$ rows of $\mathbf{H}$ and ignore the remaining. We will use $\mathbf{H}_a$ to denote the $N_a \times N_t$ truncated matrix. The dimension of the channel matrix then reduces to $2N_aN_t$, which remains a large number in the massive MIMO regime. For classical CS-based methods, $\mathbf{H}_a$ is sparse enough when $N_t \rightarrow \infty$, in other words, $\mathbf{H}_a$ does not meet the sparsity requirement with the limited $N_t$.

We are interested in designing an encoder:
\begin{equation}
\label{eq:encoder}
\mathbf{s} = F_{enc}(\mathbf{H_a}),    
\end{equation}
which can transform the channel matrix into an $M$-dimensional vector (codeword), where $M < N$ and $N = 2N_aN_t$. In this case, we can define the data compression ratio, $\gamma = M/N$. In addition, we have to design the inverse transformation (decoder) from the codeword to the original channel matrix such that: 
\begin{equation}
\label{eq:decoder}
    \hat{\mathbf{H}}_a = F_{dec}(\mathbf{s}).
\end{equation}
The CSI feedback approach works as follows. Once the channel matrix $\mathbf{\Tilde{H}}$ is acquired at the UE side, we perform 2D DFT in (\ref{eq:discr_FT}) to obtain the truncated matrix $\mathbf{H}_a$ and then use the encoder (\ref{eq:encoder}) to generate a codeword $\mathbf{s}$. The generated code word, $\mathbf{s}$, is sent to the BS over an AWGN control channel. The BS receives a noisy version of the codeword denoted by $\hat{\mathbf{s}}$ such that:
\begin{equation}
    \hat{\mathbf{s}} = \mathbf{s} + \mathbf{z},
\end{equation}
where $\mathbf{z}$ is a noise vector sampled from a standard Gaussian. Then the BS uses the decoder (\ref{eq:decoder}) to obtain an approximation for the truncated channel matrix $\mathbf{\hat{H}}_a$. The final channel matrix in the spatial-frequency domain can be obtained by performing an inverse DFT as depicted in Fig. \ref{fig:system_fig}.
\section{Proposed PRVNet for CSI Feedback}
\label{sec:proposed_model}

In the following sections, we refer to a set of CSI channel matrices as a dataset $X$ consisting of $C$ different CSI matrices indexed by $c \in \{ 1, 2, \dots C \}$. 

\subsection{Variational Autoencoders (VAE)}
\label{subsec:VAE}

The VAE consists of two models, namely encoder and decoder models. These models are trained jointly to maximize the standard VAE objective in (\ref{eq:stdr_VAE_loss}).

\begin{equation}
\small
\label{eq:stdr_VAE_loss}
     \mathcal{L}(x,\phi, \theta) = \mathbb{E}_{z\sim{q_(z|x)}} \left [log \: p_\theta(x|z) \right ] - KL(q_\phi(z|x) || p(z)),
\end{equation}

\noindent where $x$ is the input, $z$ is the latent code, $\phi$ and $\theta$ are the encoder and decoder parameters, respectively. The output of the encoder model, also known as inference model, is given by:
\begin{equation}
\label{eq:VAE_encoder}
    f_\phi (x_c) \equiv [ \mu_\phi(x_c), \sigma_\phi(x_c)] \in \mathbb{R}^{2K}
\end{equation}
where the non-linear function $f_\phi(\cdot)$ is a neural network with parameters $\phi$. Both $\mu_\phi(x_c)$ and $\sigma_\phi(x_c)$ are K-dimensional vectors representing the mean and variance of a Gaussian distribution. The latent representation, code word, $\mathbf{z}_c$ is a K-dimensional vector sampled from this distribution such that: 
\begin{equation}
\label{eq:distribution}
   q_\phi (z_c|x_c) = \mathcal{N}(\mu_\phi(x_c), diag \{ \sigma_\phi^2(x_c) \} ).  
\end{equation}

\noindent That is, for each data-point, $x_c$, in the dataset, the inference model outputs the corresponding variational parameters of a variational distribution, $q_\phi (z_c|x_c)$. When optimized, this distribution approximates the intractable posterior $p(z_c|x_c)$.

The decoder model, also known as generative model, takes the sampled codeword as input. It uses this codeword, $\mathbf{z}_c$, to reconstruct the original input, $x_c$ using a nonlinear function $p_\theta(x_c|z_c)$. The model then trained to maximize the function given by (\ref{eq:stdr_VAE_loss}) in an end-to-end fashion. 

\noindent The first term in (\ref{eq:stdr_VAE_loss}) represents the reconstruction loss between the original input and its reconstructed image. While the second term represents the KL divergence between the encoder’s distribution, $q_\phi(z|x)$, and the true distribution, $p(z)$. This divergence measures how much information is lost when using $q$ to represent a prior over $\mathbf{z}$ and encourages its values to follow a Gaussian distribution. 
Since the function in (\ref{eq:stdr_VAE_loss}) is a lower bound for the log marginal likelihood, it is referred to as the evidence lower bound (ELBO) function. We can note that ELBO is a function in both $\phi$ and $\theta$.

\subsubsection{Taxonomy of Autoencoders}

Variational autoencoders are generative models that learn a latent representation for the input data. Unlike classic autoencoders which employ a deterministic latent space (i.e., estimating a point for each dimension in the latent space), VAE employs a stochastic latent space (i.e., samples form a tractable distribution usually assumed to be a Gaussian distribution) \cite{hussien2021fault}. 

Maximum-likelihood estimation in a regular autoencoder takes the following form:

\begin{equation}
\label{eq:classic}
\begin{matrix}
    \theta ^{AE}, \phi^{AE} = {\arg \max}_{\theta, \phi} \sum_{c}^{}\mathbb{E}_\delta (z_c - g_\phi(x_c))\left [ log(p_\theta)(x_c|z_c) \right ] \\ \\
    = {\arg \max}_{\theta, \phi} \sum_{c}^{} log(p_\theta)(x_c|g_\phi(x_c))
\end{matrix}
\end{equation}

\begin{figure}[t]
\centerline{\includegraphics[width=0.45\textwidth]{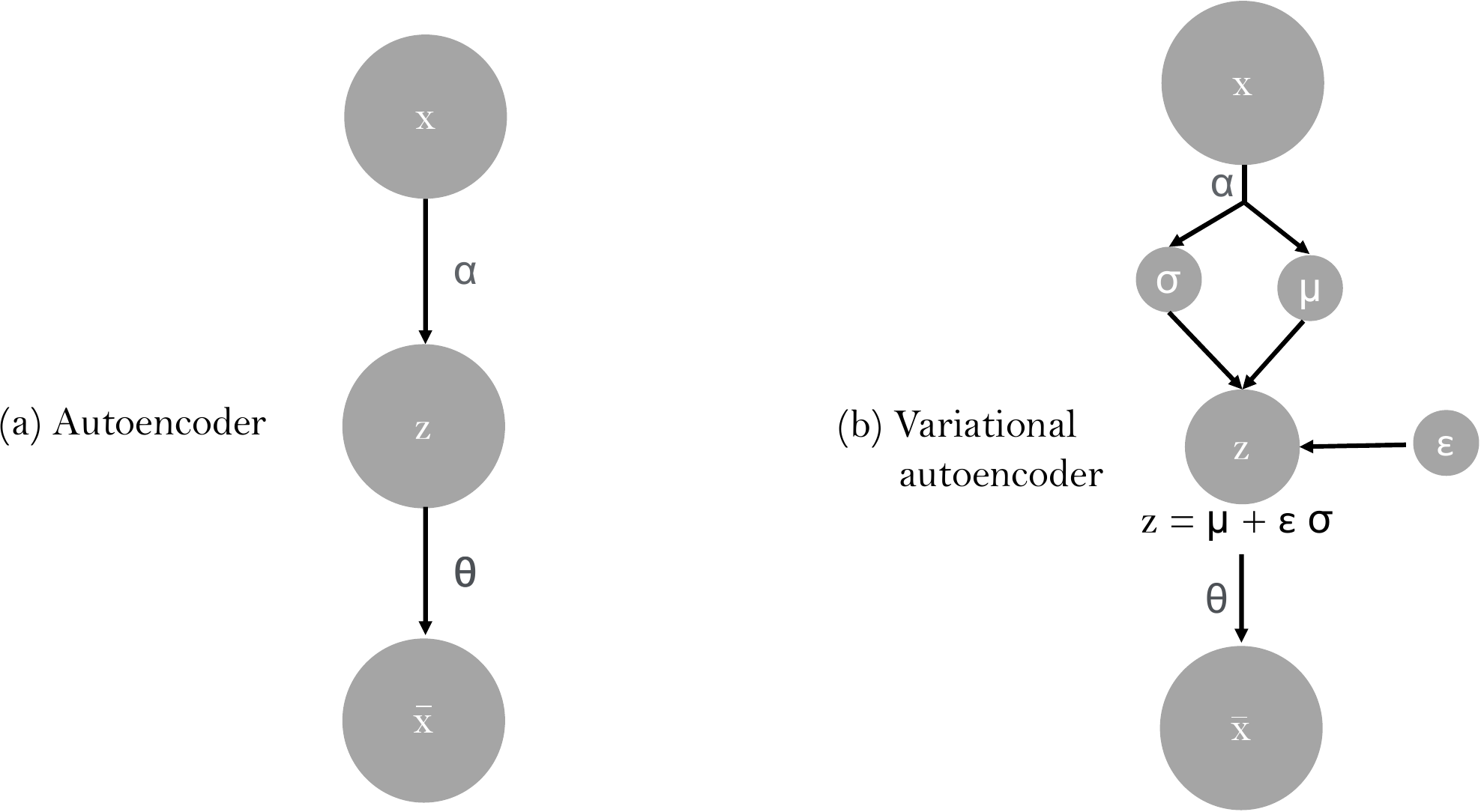}}
\caption{Variatinal autoencoder with reparameterizatino trick versus classical point estimate autoencoders.}
\label{fig:rep_trick}
\end{figure}

We can note from (\ref{eq:classic}) that classical autoencoder effectively optimizes the first term in the VAE objective using a delta variational distribution. This means that $q_\phi(z_c|x_c) = \delta (z_c - g_\phi(x_c)),$ and hence it does not regularize $q_\phi(z_c|x_c)$ toward any distribution like VAE. We can also note that $\delta (z_c - g_\phi(x_c))$ is a delta distribution with mass only at the output $g_\phi(x_c)$. Contrast this to what happens in VAE, where the learning is done using a variational distribution (i.e., $g_\phi(x_c)$ generates the parameters of a certain tractable distribution, the mean and variance in the case of Gaussian distribution). This implies that VAE has the ability to capture per-data-point variances in the latent space, $\mathbf{z_c}$. One of the main issues of autoencoders is the high possibility of overfitting which is due to the fact that the network learns to put all the probability mass to the non-zero entries in $x_c$. By introducing dropout \cite{srivastava2014dropout} at the input layer, the classical autoencoder is less prone to overfitting. Fig. \ref{fig:rep_trick} shows the main difference between point estimate autoencoders and VAE. 

\begin{algorithm}[t]
\SetAlgoLined
 Randomly initialize $\theta$ and $\phi$\;
 \While{not convergeed}{
  Sample a batch of CSI channels $\mathcal{B}$\;
  
  \ForAll{$c \in \mathcal{B}$}{
  Sample $\epsilon \in \mathcal{N}(0,I)$\;
  Compute $z_c$ using the reparamterization trick\;
  Compute noisy gradient $\triangledown_\theta \mathcal{L}$ and $\triangledown_\phi \mathcal{L}$ using the sampled $z_c$\;
  }
  Average noisy gradient for a batch\;
  Update $\theta$ and $\phi$ using stochastic gradient descent\;
 }
 Return $\theta$ and $\phi$
 \caption{VAE-SGD Training PRVNet for CSI feedback with stochastic gradient descent.}
 \label{algo:training VAE}
\end{algorithm}

\subsection{The proposed model (PRVNet)}
\label{subsec:proposed_model}

As discussed in subsection \ref{subsec:VAE}, the second term in the loss function (\ref{eq:stdr_VAE_loss}) introduces a compromise between how close the approximate posterior stays to the prior during learning and our ability to reconstruct the original data from the codeword. Therefore, we introduce a new hyperparameter $\beta$, where $\beta \neq 1$. Note that using this hyperparameter, we are no longer optimizing a lower bound on the log marginal likelihood.

Setting $\beta < 1$ means that we force the model to learn better data reconstruction and pays less attention to the prior constraint $\frac{1}{C}\sum_{c=0}^{C} q(z|x_c) \approx p(z) \approx \mathcal{N}(z;0,I_K)$. In other words, a model trained with $\beta <1$ will be less able to generate novel CSI matrices by ancestral sampling. On the other hand, setting $\beta >1$ emphasizes the importance of the prior distribution constraint over the ability to reconstruct the input from the codeword. Note that setting $\beta$ to zero eliminates the prior distribution constraint and reduces the loss function to that of the classical point estimate autoencoders.

Recall that our goal is to make a good reconstruction at the BS side without generating novel imagined CSI matrices. Treating $\beta$ as a free hyperparameter, with $\beta <1$, therefore can significantly improve the reconstruction results without any additional cost in terms of time or the number of model parameters. Therefore, we propose objective function in (\ref{eq:PRVNet}). Since we can interpret the second term as a regularization term, we coin a model trained with (\ref{eq:PRVNet}) by partially regularized VAE network (PRVNet).

\begin{equation}
\label{eq:PRVNet}
     - \mathbb{E}_{z\sim{q_(z|x)}} \left [log \: p_\theta(x|z) \right ] + \beta \cdot KL(q_\phi(z|x) || p(z))
\end{equation}

\subsubsection{Selecting a value for $\beta$}
We propose an algorithm for selecting the best value of $\beta$. At the beginning of the training phase, we set $\beta=0$ and gradually increase its value to 1. We linearly anneal the KL term slowly over a large number of gradient updates to $\phi$ and  $\theta$ and record the best value of $\beta$ when the performance reaches the peak \cite{liang2018variational}. After figuring out the best value of $\beta$, which we denote here as $\beta_*$, we retrain the model with the values of $\beta$ starting from 0 to $\beta_*$. If the computation power is limited, we can stop increasing $\beta$ once we notice a degradation in the validation metric. In this way, training our model does not incur any additional cost compared with training traditional VAE models.  

\subsubsection{Training PRVNet}
\label{sub:training_PRVNet}

Recall that the proposed model optimizes the function in (\ref{eq:PRVNet}) while VAE is trained to optimize the standard ELBO function given in (\ref{eq:stdr_VAE_loss}). We can obtain an unbiased estimate of (\ref{eq:PRVNet}) by sampling $\mathbf{z_c} \sim{q_\phi}$ and optimize it by stochastic gradient descent. However, the challenge is that we cannot trivially take gradients with respect to $\phi$ through this sampling. The reparameterization trick solves this challenge by sampling $\epsilon \sim{\mathcal{N}(0, I_k)}$ and reparametrize the generated latent code such that, $\mathbf{z_c} = \mu_\phi (x_c) + \epsilon \odot \sigma_\phi(x_c)$ \cite{kingma2013auto}. This way, the stochasticity in the sampling process is eliminated and the gradient with respect to $\phi$ now can be back-propagated through the sampled latent code $\mathbf{z_c}$. A detailed description for the training process is given in Algorithm \ref{algo:training VAE}.

\section{Simulation Results and Analysis}
\label{sec:sim_results}

\subsection{Experiment Setup}
\label{subsec:exper_setup}

\paragraph{Architecture Details}
The encoder and decoder models are convolutional neural networks (CNN)-based architectures. We set the batch size to $128$. The model weights are initialized according to \textit{He} initialization \cite{he2015delving}. We optimize the model using \textit{Adam} optimizer \cite{kingma2014adam} with $0.1$ learning rate for $1000$ epochs. The function proposed in (\ref{eq:PRVNet}) is used as the model loss function. To alleviate the effect of overfitting, we employ a weight decay of $1^{-4}$ for kernel and bias weights.

\paragraph{Dataset} We consider two types of scenarios as given in \cite{wen2018deep}: the outdoor scenario at 300MHz and the indoor scenario at 5.3GHz. The channels are generated following the default settings of COST 2100 \cite{liu2012cost}. At the BS, a uniform linear array (ULA) with $N_t=32$ is considered. For the FDD system, we set $N_c=1024$ in frequency domain and $N_a=32$ in angular domain. The dataset contains $150,000$ independently generated channels divided into three parts. The train, validation, and testing parts consist of $100,000$, $30,000$ and $20,000$ channel matrices, respectively.

\subsection{Performance of PRVNet}
\label{subsec:perform_PRVNet}

\begin{table}[t]

\caption{Comparison of NMSE (db) for different methods.}

\begin{tabular}{p{1.0cm}|p{3cm}|p{1.5cm}|p{1.5cm}}
\hline \hline
\multicolumn{1}{c|}{\multirow{2}{*}{\textbf{CR}}} & \multicolumn{1}{c|}{\multirow{2}{*}{Methods}} & \multicolumn{2}{c}{\textbf{NMSE (db)}}                                      \\ \cline{3-4} 
\multicolumn{1}{c|}{}                             & \multicolumn{1}{c|}{}                         & \multicolumn{1}{l|}{\textbf{Indoor}} & \multicolumn{1}{l}{\textbf{Outdoor}} \\ \hline \hline
\multirow{6}{*}{1/4}                               & LASSO                                         & -7.59                                & -5.08                                 \\ \cline{2-4} 
                                                   & BM3D-AMP                                      & -4.33                                & -1.33                                 \\ \cline{2-4} 
                                                   & TVAL3                                         & -14.87                               & -6.90                                 \\ \cline{2-4} 
                                                   & CsiNet                                        & -17.36                               & -8.75                                 \\ \cline{2-4} 
                                                   & CRNet                                         & -26.99                               & -12.71                                \\ \cline{2-4} 
                                                   & PRVNet (our)                                  &   \textbf{-27.7}                      &     \textbf{-13.9}                                  \\ \hline \hline
\multirow{6}{*}{1/16}                              & LASSO                                         & -2.72                                & -1.01                                 \\ \cline{2-4} 
                                                   & BM3D-AMP                                      & 0.26                                 & 0.55                                  \\ \cline{2-4} 
                                                   & TVAL3                                         & -2.61                                & -0.43                                 \\ \cline{2-4} 
                                                   & CsiNet                                        & -8.65                                & -4.51                                 \\ \cline{2-4} 
                                                   & CRNet                                         & -11.35                               & -5.44                                 \\ \cline{2-4} 
                                                   & PRVNet (our)                                  &  \textbf{-13}                          & \textbf{ -6.1}                                     \\ \hline \hline
\multirow{6}{*}{1/32}                              & LASSO                                         & -1.03                                & -0.24                                 \\ \cline{2-4} 
                                                   & BM3D-AMP                                      & 24.72                                & 22.66                                 \\ \cline{2-4} 
                                                   & TVAL3                                         & -0.27                                & 0.46                                  \\ \cline{2-4} 
                                                   & CsiNet                                        & -6.24                                & -2.81                                 \\ \cline{2-4} 
                                                   & CRNet                                         & -8.93                                & -3.51                                 \\ \cline{2-4} 
                                                   & PRVNet (our)                                  &  \textbf{-9.52}                        & \textbf{-4.23}                                       \\ \hline \hline
\multirow{6}{*}{1/64}                              & LASSO                                         & -0.14                                & -0.06                                 \\ \cline{2-4} 
                                                   & BM3D-AMP                                      & 0.22                                 & 25.45                                 \\ \cline{2-4} 
                                                   & TVAL3                                         & 0.63                                 & 0.76                                  \\ \cline{2-4} 
                                                   & CsiNet                                        & -5.84                                & -1.93                                 \\ \cline{2-4} 
                                                   & CRNet                                         & -6.49                                & -2.22                                 \\ \cline{2-4} 
                                                   & PRVNet (our)                                  & \textbf{-6.9}                                      & \textbf{-2.53}                                      \\ \hline \hline
\end{tabular}
\label{table:NMSE}
\end{table}

We compare the performance of PRVNet with three CS-based methods, namely, Lasso $L_1$-solver \cite{daubechies2004iterative}, TVAL-3 \cite{chengbo2009tv}, and BM3D-AMP \cite{metzler2016denoising}. Moreover, two recent deep learning-based methods, namely, CsiNet \cite{wen2018deep} and CRNet \cite{lu2020multi}, are also considered in the comparison. To evaluate the performance of different methods, we measure the distance between the original CSI matrix, $\mathbf{H}_a$, and the reconstruction image, $\mathbf{\hat{H}}_a$, using the normalized mean square error \eqref{eq:NMSE}.

\begin{equation}
\textnormal{NMSE (db)} = 10\log \; \mathbb{E}\left(  \frac{{\left\| \mathbf{H}_a - \mathbf{\hat{H}}_a \right\|}_2^{2}}{{\left\| \mathbf{H}_a \right\|}_2^{2}}  \right).
\label{eq:NMSE}
\end{equation}

\begin{figure}[t]
	\centering 
	\includegraphics[width=0.49\textwidth]{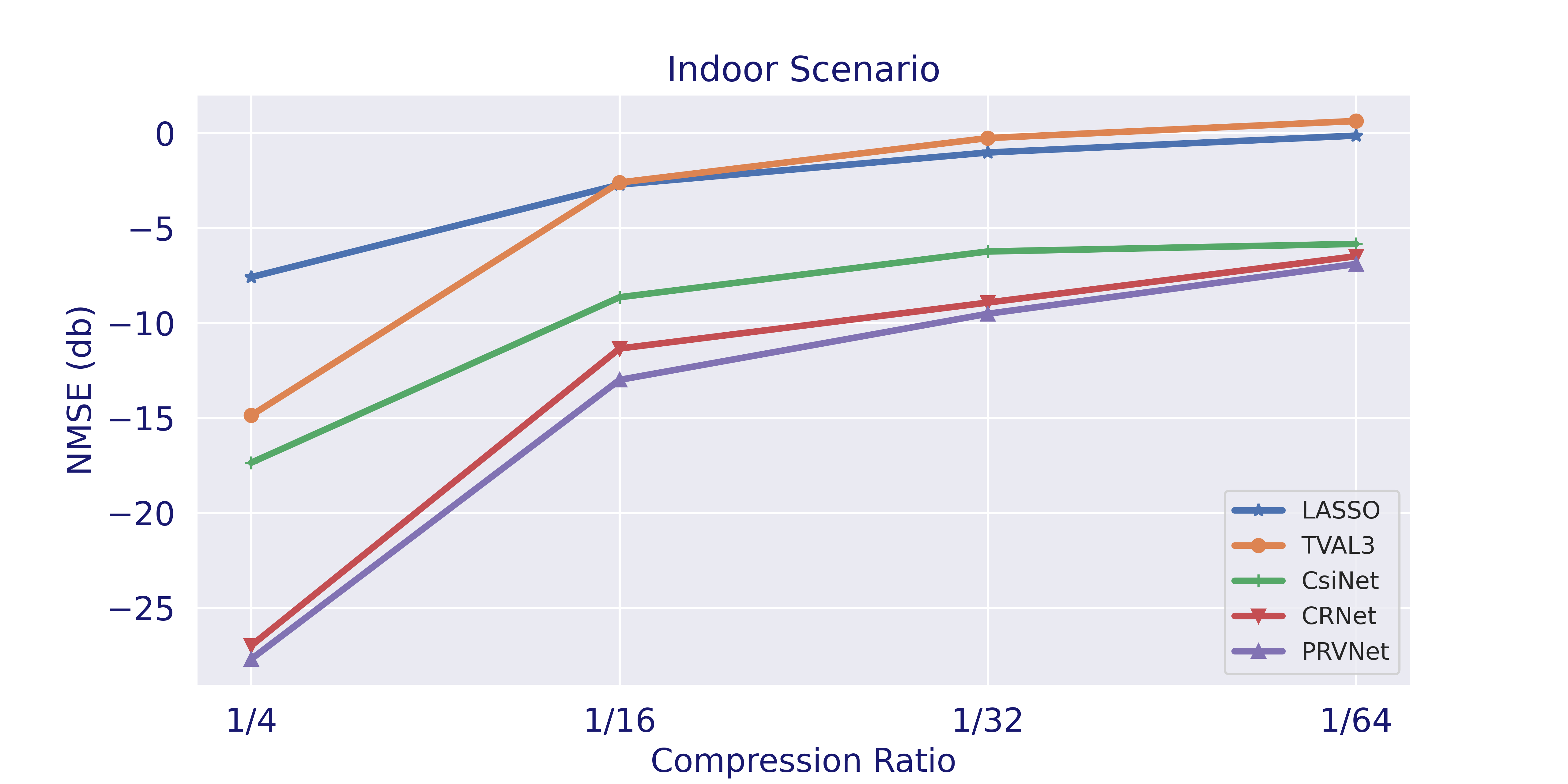}
	\includegraphics[width=0.49\textwidth]{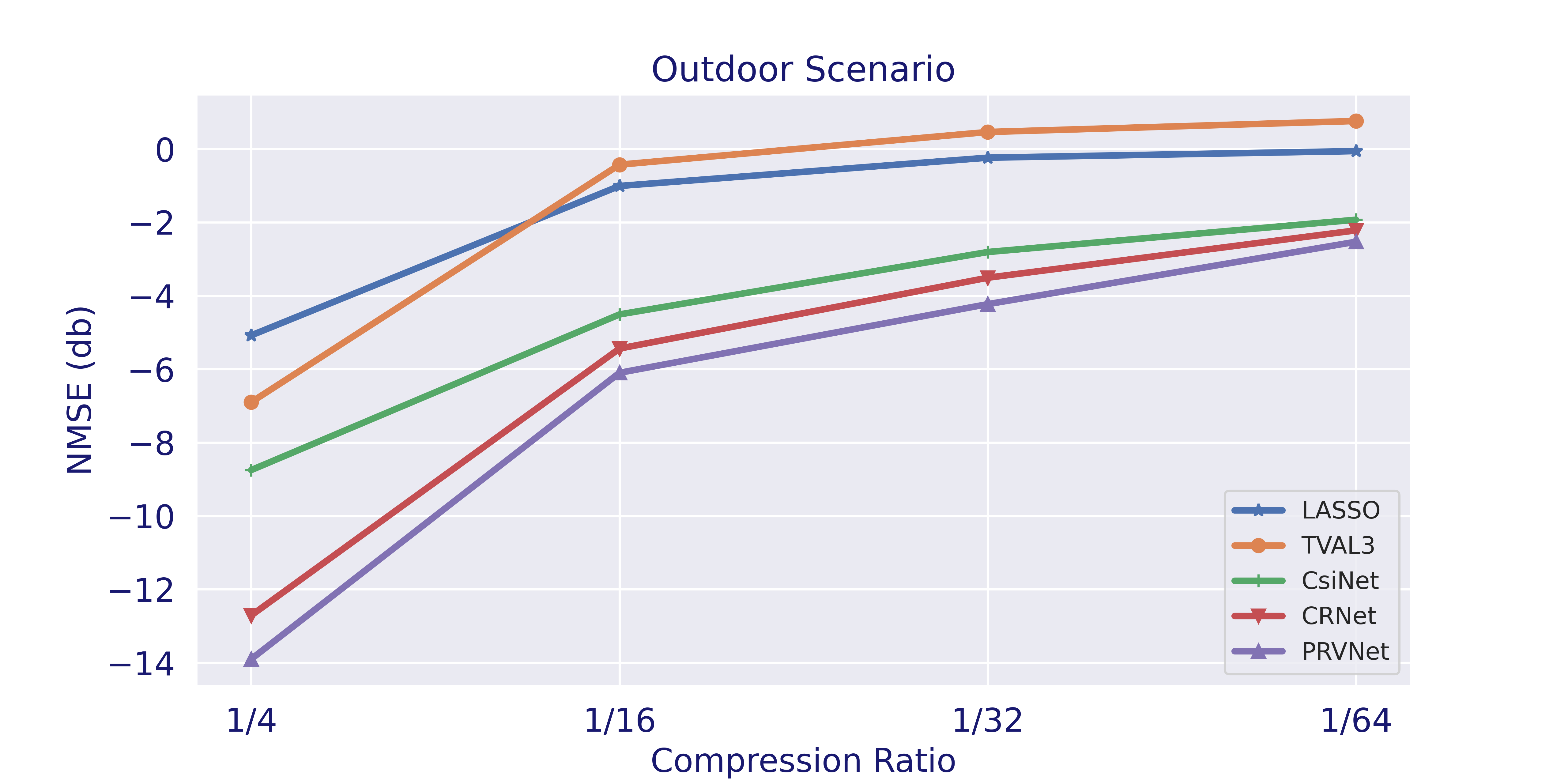}
	\caption{Top: indoor, bottom: outdoor. Comparison (in terms of the reconstruction loss measured in NMSE) between the proposed model and other works in literature.}
	\label{fig:NMSE_Comparison} 
\end{figure}

Table \ref{table:NMSE} and Fig. \ref{fig:NMSE_Comparison} show the performance of the proposed PRVNet against different benchmark models. We can see that PRVNet outperforms all classical CS-based methods as well as recent deep learning-based methods. PRVNet with the proposed loss function in \eqref{eq:PRVNet} is capable of capturing CSI features to increase the reconstruction accuracy at the BS. Unlike the other benchmark models which do not consider the channel noise in their model design, an advantage of the proposed model is its robustness against different noise levels. This property will be further investigated shortly. 

\begin{table}[t]

\caption{The effect of annealing $\beta$ to different values on indoor scenario for a compression ratio 1/4.}

\begin{tabular}{p{5.0cm}|p{2.0cm}}
\hline \hline
\hfil $\beta$ annealing strategy                & \hfil NMSE (db) \\ \hline \hline
\hfil No annealing                           & \hfil -25.83    \\ \hline
\hfil Annealing $\beta$ to the maximum ($\beta$=1) & \hfil -26.32    \\ \hline
\hfil Annealing $\beta$ to 0.3                  & \hfil \textbf{-27.7}     \\ \hline
\end{tabular}
\label{table:beta}
\end{table}

The effect of $\beta$ annealing has been studied and demonstrated in Table \ref{table:beta}. We can see that the model achieved the highest NMSE when no $\beta$-annealing has been applied. Under the same dataset and compression ratio, the model achieved lower NMSE when $\beta$ has been annealed to 1. The best NMSE has been achieved with annealing $\beta$ from 0 to 0.3 and complete the training epochs without further increase in the value of $\beta$. Although, this value might be sub-optimal compared to a thorough grid search. The proposed algorithm is much more efficient, and achieves a similar performance.

\begin{figure}[t]
\centerline{\includegraphics[width=0.49\textwidth]{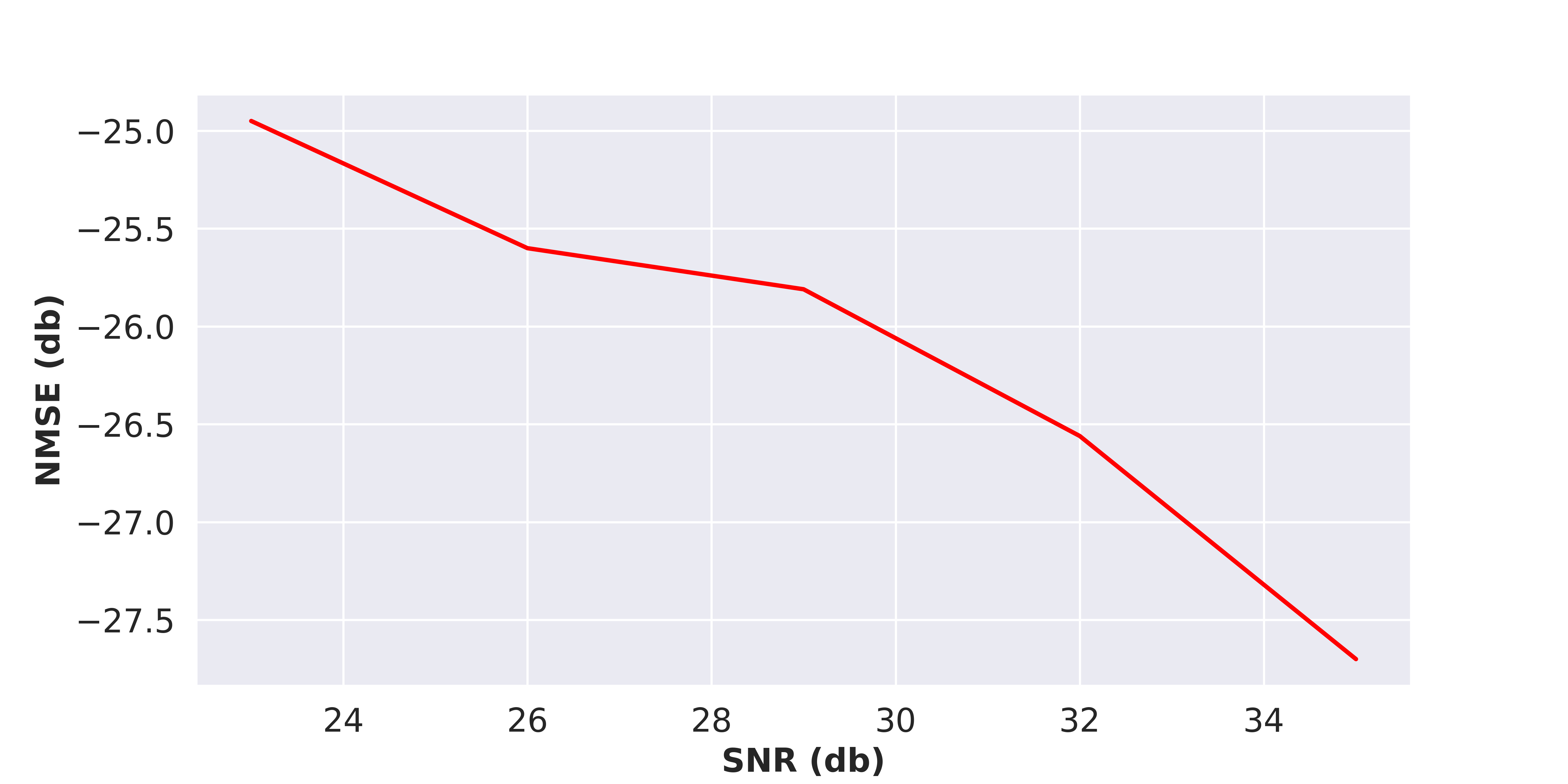}}
\caption{The NMSE(db) under different SNR(db) values.}
\label{fig:NMSE_SNR}
\end{figure}

\begin{table}[t]
\centering
\caption{The robustness of the proposed PRVNet under different signal-to-noise ratios.}
\begin{tabular}{p{3.0cm}|p{2.0cm}}
\hline \hline
\hfil SNR (db)  & \hfil NMSE (db) \\ \hline \hline
\hfil 35        & \hfil -27.7     \\ \hline
\hfil 32        & \hfil -26.56    \\ \hline
\hfil 29        & \hfil -25.81    \\ \hline
\hfil 26        & \hfil -25.6     \\ \hline
\hfil 23        & \hfil -24.95    \\ \hline
\end{tabular}
\label{table:noise}
\end{table}

To further evaluate the robustness of the proposed model under different noise conditions, we simulate the AWGN feedback channel by adding a random Gaussian noise to the codeword and pass it through the decoder model such that:

\begin{equation}
\label{eq:noise_ch}
    \mathbf{\bar{z}} = \mathbf{z} + \mathbf{\epsilon},
\end{equation}

\noindent where  $\epsilon \sim{\mathcal{N}(0, \sigma_n)}$. The degradation of the NMSE with different SNR values is shown in Table \ref{table:noise}. We can see that the proposed PRVNet model shows an outstanding robustness against different noise-levels, see Fig. \ref{fig:NMSE_SNR}. We notice a slow degradation in the NMSE when the noise level increases, which indicates that the codewords generated by the proposed PRVNet model can still convey relevant features about the original CSI matrix even under noisy conditions. This can be explained by the fact that PRVNet, unlike other models in the literature, learns a distribution for each dimension in the codeword. This makes the effect of the noise much less severe than in point estimate models because even with noise, a value in the codeword may still look as being sampled from the same learned distribution.

\section{Conclusion}
In this paper, a novel deep learning model, PRVNet, has been proposed for downlink channel state information (CSI) feedback in MIMO-FDD systems. The PRVNet customizes the traditional variational autoencoder objective to incorporate the special characteristics of CSI feedback problem. Unlike prior work that assume ideal feedback channel, we modeled an AWGN feedback channel and proved that the codewords generated by PRVNet are more robust against varying noise conditions. The proposed model outperforms state-of-the-art deep learning-based and compressive-sensing based models in both noise-free and noisy channel conditions.

\section*{Acknowledgment}
The authors thank Mitacs and Ciena for supporting this research in the IT13947 grant.

\bibliographystyle{./bibliography/IEEEtran}
\bibliography{./bibliography/IEEEabrv,./bibliography/References}

\end{document}